\documentclass[reprint, superscriptaddress, amsmath,prl,amssymb, aps, floatfix]{revtex4-2}

\makeatletter
\newsavebox{\@brx}
\newcommand{\llangle}[1][]{\savebox{\@brx}{\(\m@th{#1\langle}\)}%
  \mathopen{\copy\@brx\kern-0.5\wd\@brx\usebox{\@brx}}}
\newcommand{\rrangle}[1][]{\savebox{\@brx}{\(\m@th{#1\rangle}\)}%
  \mathclose{\copy\@brx\kern-0.5\wd\@brx\usebox{\@brx}}}
\usepackage{lipsum}
\usepackage{graphicx}
\usepackage{dcolumn}
\usepackage{bm}
\usepackage{xcolor}
\usepackage{xspace}
\usepackage{comment}

\usepackage[colorlinks,allcolors=blue]{hyperref}
\usepackage[makeroom]{cancel}
\usepackage{float}
\usepackage{siunitx}
\usepackage{color}
\definecolor{dgreen}{rgb}{0,0.7,0}
\def\bluew#1{{\color{red} #1}}

\def\bluew#1{{\color{black} #1}}
\def\blueww#1{{\color{black} #1}}
\newcommand{\beq}{\begin{equation}}
\newcommand{\eeq}{\end{equation}}
\newcommand{\bea}{\begin{eqnarray}}
\newcommand{\eea}{\end{eqnarray}}

\usepackage{subfigure}
\usepackage{booktabs} 
\usepackage{natbib}
\usepackage[normalem]{ulem}

\begin{document}

\title{\blueww{Sokoban Random Walk: From Environment Reshaping to Trapping Crossover}
}
\author{Prashant Singh} 
\email{prashantsinghramitay@gmail.com}
\author{David A. Kessler}
\address{Department of Physics, Bar-Ilan University, Ramat Gan 52900, Israel}
\author{Eli Barkai} 
\address{Department of Physics, Bar-Ilan University, Ramat Gan 52900, Israel}
\address{Institute of Nanotechnology and Advanced Materials, Bar-Ilan University, Ramat Gan 52900, Israel}

\vspace{10pt}

\begin{abstract}

We study the dynamics of a \textit{Sokoban random walker} moving in a disordered medium with obstacle density $\rho$. In contrast to the classic model of de Gennes with static obstacles that exhibits a percolation transition, the Sokoban walker is capable of modifying its environment by pushing a few surrounding obstacles. Surprisingly, even a limited pushing ability leads to a loss of the percolation transition. Through a combination of a rigorous large-deviation calculation and extensive numerical simulations, we demonstrate that the Sokoban model belongs to the Balagurov-Vaks-Donsker-Varadhan trapping universality class. The survival probability that the walker has not yet been trapped inside a cage exhibits stretched-exponential relaxation at late times. \blueww{Furthermore, using the average trap size as a proxy, we identify the emergence of a dynamical crossover at a density $\rho_* \approx 0.55$ between two qualitatively different trapping mechanisms: a self-trapping mechanism at low density, where the walker becomes dynamically localized within a self-formed trap, and a pre-existing trapping mechanism at high density, where confinement arises from the initial arrangement of obstacles. This crossover is responsible for the loss of the classical percolation transition.}

\end{abstract}

\maketitle

\noindent
\textit{Introduction:} The ability of a moving tracer to alter its surroundings is a natural feature of many real-world systems. From energized active particles that push surrounding obstacles \cite{expt-2, expt-3} to navigating robots modifying their terrain \cite{expt-1, Vatash2025Resetting} and kinesin proteins buckling the microtubule \cite{KABIR2020249}, such interactions between the tracer and the medium are common and can profoundly influence its transport properties. \bluew{For example, a navigating bristle robot, moving in an arena of mobile obstacles, can convert the vibrational kinetic energy into directed motion, and its collisions with movable obstacles allow it to modify the surrounding environment \cite{expt-1}}. Similarly, in biological systems, migrating immune cells can chemically remodel their environment, thereby facilitating their own migration \cite{Friedl2003}. \\
\indent In these examples, moving tracers interact with their surroundings in ways that are constrained by a finite energy resource, size, or force. As a result, their ability to modify the environment is often limited. For example, a migrating cell may modify only a portion of the extracellular matrix, or a robot may displace only a limited number of surrounding obstacles. This finiteness of the capacity to alter the environment introduces a natural constraint on tracer-environment interactions. Unlike traditional models, where the environment is either completely fixed \cite{degenne, Stauuffer,Havlin-1, Havlin-2, Weiss-1, neww2005,neww2014, neww2015, neww2015d, neww2023, neww2023d, DDhar2024} or globally time-evolving \cite{Rednerbook,randommove2, randommove3,randommove1,randommove4,randommove5, VDannea}, very little is known about systems where a tracer can dynamically reshape its local environment \cite{Shlomi-1, Shlomi-2}.\\
\indent To grasp some fundamental features of environmental reshaping on tracer dynamics, Reuveni and coworkers recently introduced a new type of model called the \textit{Sokoban random walk} \cite{Shlomi-1}. In this model, a random walker moves on a lattice in which obstacles are initially randomly distributed with density $\rho$ (with $0 \leq \rho \leq 1$). Moreover, the walker possesses a limited ability to push certain blocking obstacles, thereby inducing minimal modifications to its environment. In the absence of any pushing mechanism, this model reduces to the celebrated \textit{Ant in a labyrinth} model introduced by de Gennes in percolation \cite{degenne, Stauuffer,Havlin-1, Havlin-2, Weiss-1, neww2005,neww2014, neww2015, neww2015d, neww2023, neww2023d, DDhar2024}. For obstacle densities below a critical threshold, $\rho \leq \rho_c$ (with $\rho _c = 0$ in $1d$ and $\rho _c \approx 0.407$ in $2d$) \cite{Stauuffer}, a spanning cluster of vacant sites emerges, inducing a long-range connectivity. A random walker placed within such a cluster, and lacking any ability to modify its environment, can eventually percolate, \emph{i.e.}, escape to infinity. Surprisingly, the study in \cite{Shlomi-1} showed that the percolation transition is lost in two dimensions. This means that at low obstacle densities, where escape is easy, the tracer's ability to modify its surroundings leads to a localization that eliminates the long-range transport. Thus, even a small departure from de Gennes' model yields a fundamentally different low-density behavior, eliminating the percolation transition. \\
\indent Questions then arise - what physical principle governs the low-density transport of the Sokoban walk? How does it dynamically relax towards the localized state induced by its pushing ability? And are these principles robust against variations in the model properties? 
In this Letter, we address these key questions by analyzing the trapping behavior of the Sokoban random walk. \blueww{We show that the pushing dynamics leads to the emergence of a dynamical crossover between two trapping mechanisms which, unlike the percolation transition, is smooth and explains the different low-density physics of the model. Quite remarkably, this crossover is completely robust against variations in the pushing strength of the walker, and disappears only for the de Gennes' model, where pushing is absent.} Our findings uncover a distinct transport regime leading to the loss of percolation, providing fundamental insights into transport in dynamically reshaped environments.


\textit{Model:} Consider a square $d$-dimensional lattice system, where every site can accommodate an obstacle with probability $\rho$ and remain vacant with the complementary probability $(1-\rho)$. We place our random walker initially at the origin. At each time step, it chooses one of its neighboring sites with an equal probability and attempts a jump to that site.  The jump is successful if the target site is vacant. If the chosen site is occupied by an obstacle, the walker can still move by pushing the obstacle one step further in the same direction. The pushing is allowed provided the next site beyond the obstacle is vacant. If this condition is met, both the walker and the obstacle move in that direction. Otherwise, they do not move. Hence, the Sokoban walker is capable of modifying its local environment by pushing the surrounding obstacles and it can at most push one obstacle. This limited pushing rule introduces a dynamical constraint: motion is allowed only when certain local configurations are satisfied. Such constraints are a hallmark of kinetically constrained models (KCMs), which are widely studied in the context of glassy systems \cite{KCMN1, KCMN2, KCM1, KCM2, KCM3}. But unlike standard KCMs, where the constraint is maintained globally, in the Sokoban model the constraints are dynamically emergent and evolve locally in the vicinity of the walker.



\textit{Reactive trapping versus trapping by caging:} Consider a walker in a quenched environment without pushing ability. A well-studied problem in this context is that of the walker becoming trapped upon its first encounter with an obstacle. \cite{Balagurov1974, Rosenstock, Donsker, Weiss1980,  Grassberger1982, Redner-1983,Kayser1983,zumofen1983, Blumen1983-2, Anlauf1984, SHavlin-numerical-1984,Lubensky1984,Weiss1d1984,SHavlin_PRL_1984, Nieuwenhuizen1989, Gallos2001, Beijren2001, newadd2, newadd3, newadd1, PSS2011, PSS2013, PSS2020, PSS2025}. We refer to this form of trapping as \textit{reactive trapping}, since the obstacles act as perfect reaction centers, and the walker is absorbed by them upon the very first contact. Clearly here, there is no percolation transition, since the walker in any dimension and any  finite density will eventually become trapped by an obstacle. Balagurov and Vaks \cite{Balagurov1974} and Donsker and Varadhan \cite{Donsker} showed that the long-time survival probability $\phi(n)$ that the walker has not yet encountered any obstacle up to time $n$ in $d$-dimensions is given by
\begin{align}
\phi (n) \sim \exp \left( - \beta _{d} \lambda ^{\frac{2}{d+2}} n^{\frac{d}{d+2}} \right),~~~~ \text{with }n \lambda ^{2/d} \gg 1,
\label{original-eqn}
\end{align}
where $\lambda = |\ln(1-\rho)|$ and $\beta _d$ is a constant whose value depends on the lattice structure as well as on dimension: On a square lattice, $\beta _ 1 =3\pi^{2/3}/2 $ and $ \beta _2 \approx 3.4$ \cite{Donsker, Rednerbook}. We refer to Eq.~\eqref{original-eqn} as the Balagurov–Vaks–Donsker–Varadhan (BVDV) theory.

\begin{figure}[]
 \textbf{(a)~~~~~~~~~~~~~~~~~~~~~~~~~~~~~~(b)} \\
   \centering
	\includegraphics[width=0.2\textwidth,height=0.16\textheight]{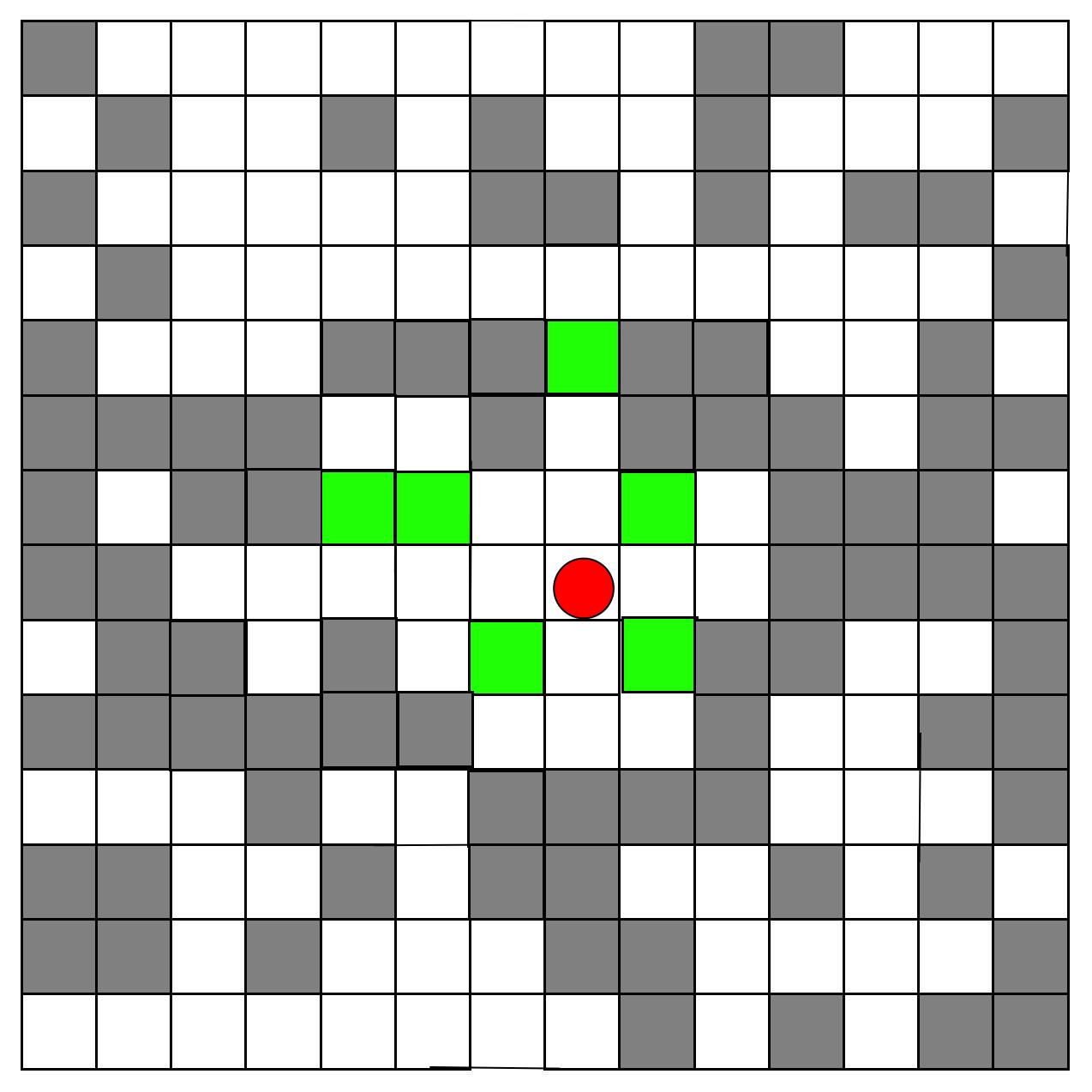}
	\hspace{0.2 cm}
	\includegraphics[width=0.2\textwidth,height=0.16\textheight]{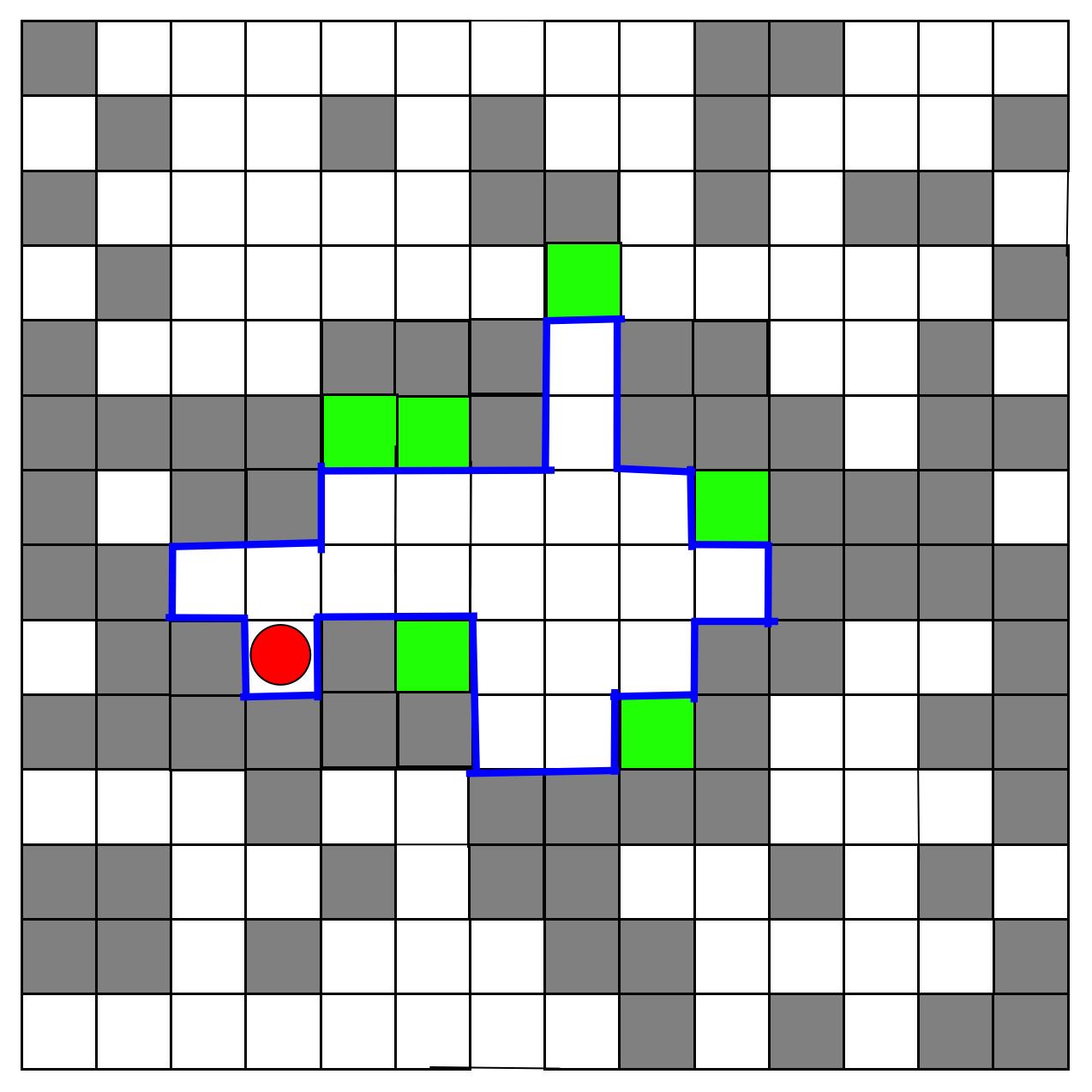}
	\caption{(a) The initial configuration of the obstacles in two dimensions with walker shown in red and the obstacles shown in gray and green. They are identical, but distinguished here to illustrate that the green ones will be pushed by the walker. (b) This represents the trapped scenario where, the green obstacles have been pushed and the walker cannot visit any new lattice site further. The trap size $A_{\rm T} = 21$ is demonstrated in blue.
	}
	\label{fig-update}
\end{figure}

To investigate the absence of the percolation transition in the Sokoban model, we consider a different form of trapping in which we monitor the number of distinct lattice sites visited by the walker. For a given realization of the motion, this number is a non-decreasing function of time, and it saturates only if the motion is caged in a trap from all directions. Such a saturation signals that the walker cannot access arbitrarily distant sites and therefore cannot percolate to infinity. For instance, Fig.~\ref{fig-update}(a) shows the initial configuration of obstacles, which gets modified by the walker to that of Fig.~\ref{fig-update}(b). At this point, the walker does not visit any new lattice site further and is confined inside a fixed domain of finite size $A_{\rm T}$. We refer to this phenomenon as \textit{trapping by caging}, and to the confining domain as a trap. Furthermore, the time $n_{\rm T}$ at which the saturation value is reached for the first time will be referred to as the trapping time.

\bluew{The Sokoban model provides a conceptual lens through which irreversible caging can be viewed. Here, the motion of a single tracer continually reorganizes the surrounding obstacles, until a final cage emerges in an irreversible manner. This notion of caging is qualitatively distinct from the transient cages commonly encountered in dense active-passive mixtures \cite{NEB-1,NEB-2, NEB-3}, where local relaxation eventually restores mobility. We will return to the wider significance of this mechanism in a later section.}

\begin{figure*}[t]
	\centering	
	\includegraphics[width=0.4\textwidth,height=0.24\textheight]{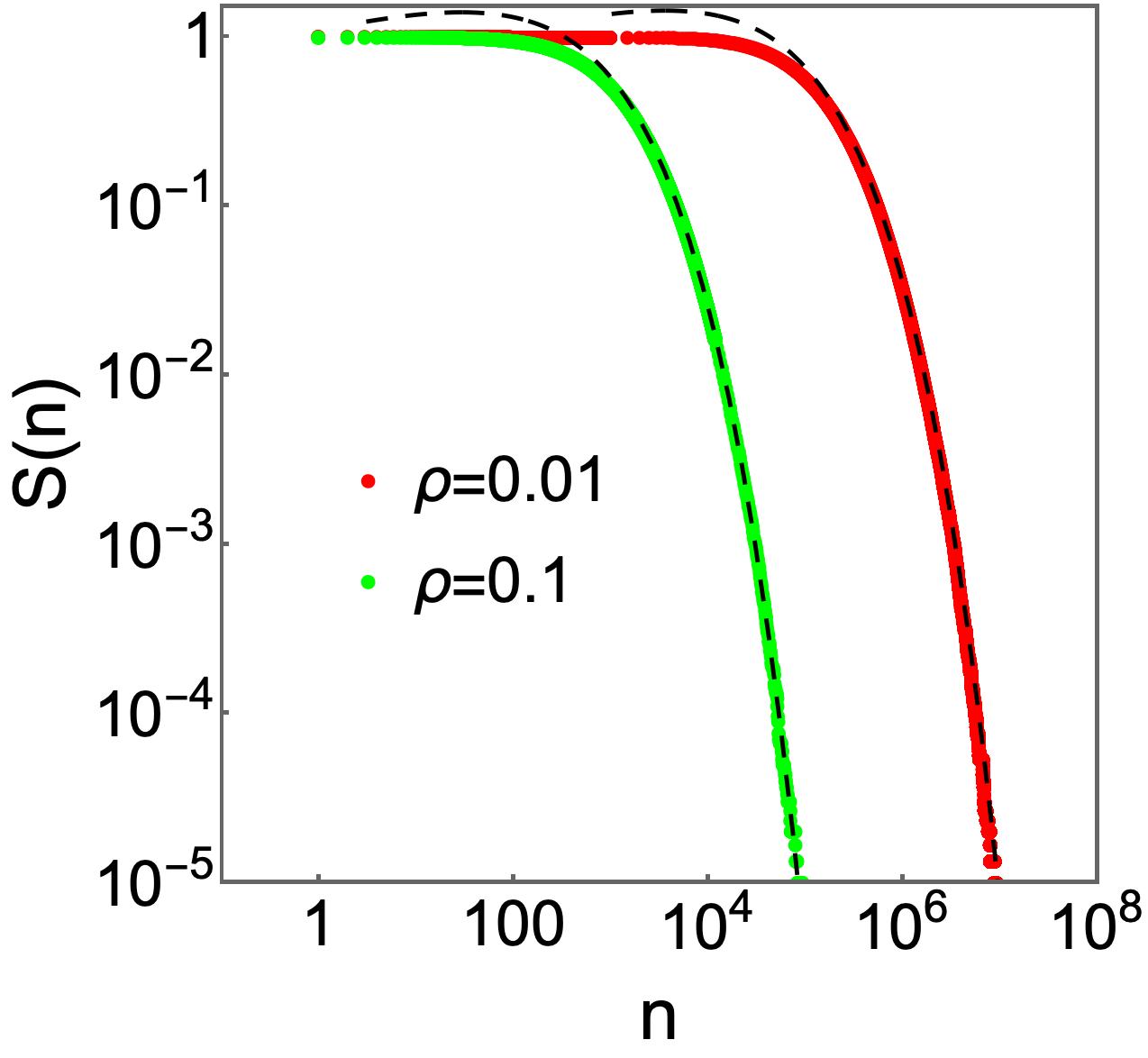} \hspace{1.cm}
	\includegraphics[width=0.4\textwidth,height=0.24\textheight]{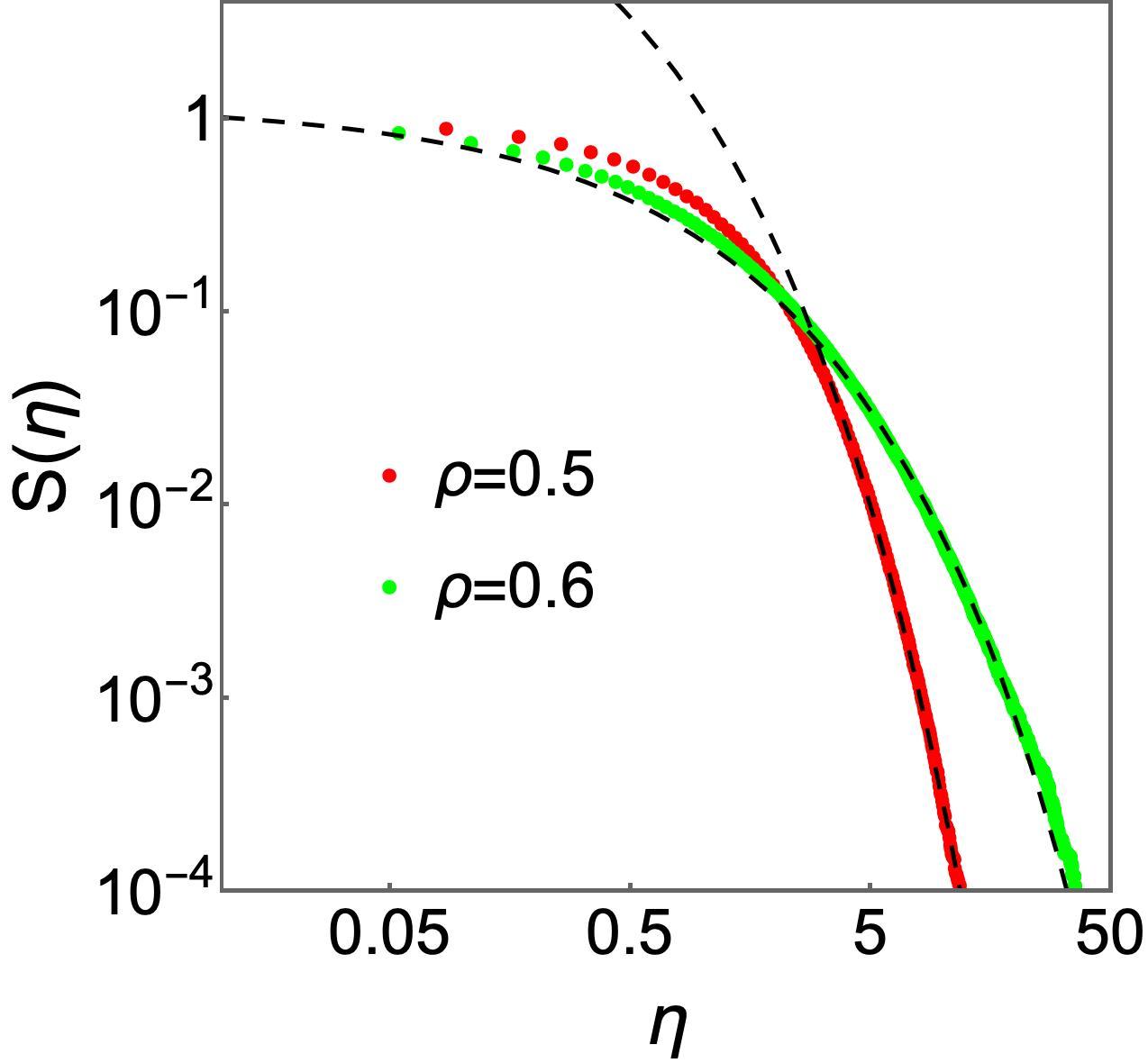}
	\caption{Long-time relaxation of the survival probability as a function of time in one (left panel) and two (right panel) dimensions. The left panel shows a comparison between numerical simulations (colored symbols) and the theoretical result given in Eq.~\eqref{sok-surv-eq-35} (dashed lines). The right panel, on the other hand, shows the survival probability as a function of $\eta = n / \langle n_{\rm T} \rangle $. Here the dashed lines are the fits to the simulation data given in Eq.~\eqref{ahbvci} (shown by symbols).	 
	}
	\label{fig-short-n-1d}
\end{figure*}

\textit{Summary of results:} A natural question is what are the statistical properties of $A_{\rm T}$ and $n_{\rm T}$? Our first result concerns the survival probability $S(n)$ that the Sokoban walker has not yet been trapped in a cage until time $n$. At late times, it exhibits stretched-exponential relaxation, with stretch exponents $1/3$ in one dimension and $1/2$ in two dimensions, see Fig.~\ref{fig-short-n-1d}. Both of them match with the BVDV formula in Eq.~\eqref{original-eqn}. \bluew{Notably, these exponents are robust under variations in the microscopic rules of the Sokoban walker and remain applicable even to generalized versions of the model. We establish this universality through a rigorous large-deviation calculation for generic pushing rules in one dimension and extensive numerical simulations in two dimensions. This robustness also makes these results amenable to experimental verification in setups such as active colloids or self-propelled robots \cite{expt-2, expt-3, expt-1}}

\begin{figure}[t]
	\centering
	\includegraphics[width=0.43\textwidth,height=0.22\textheight]{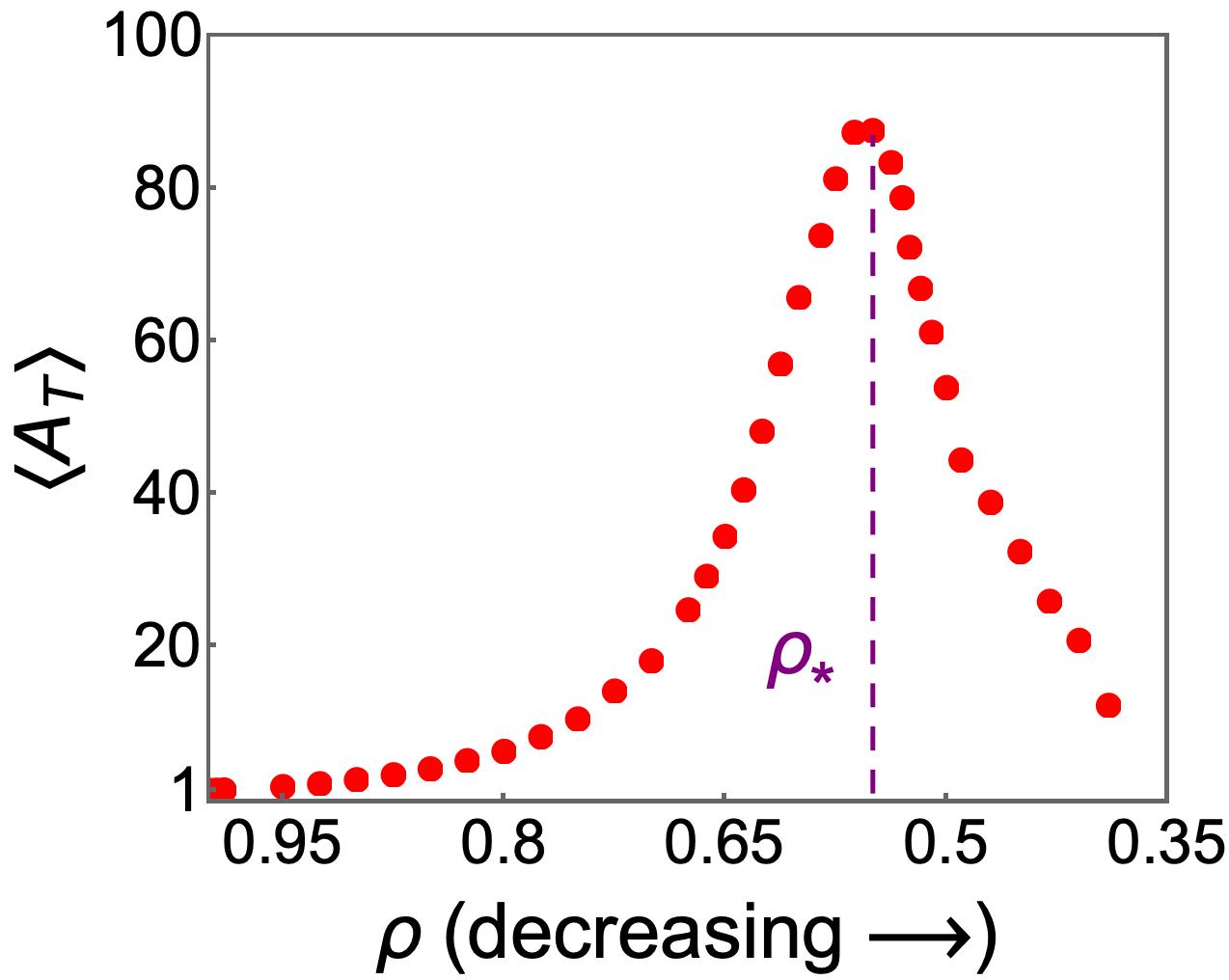}
	\caption{Nonmonotonic behavior of $\langle A_{\rm T} \rangle $ as a function of $\rho$ for a two-dimensional Sokoban walker using numerical simulations (red symbols). The turnover density is $\rho _{*} \approx 0.55$, marking a \blueww{crossover} from the low-density self-trapping to the pre-existing traps at high density. 
	}
	\label{fig-2d-area}
\end{figure}

\blueww{We next show that although the percolation transition is lost in the two-dimensional Sokoban model \cite{Shlomi-1}, it nonetheless exhibits a new dynamical crossover at a characteristic density $\rho_* \approx 0.55$ between two qualitatively different trapping mechanisms -- self-traps at low densities to the pre-existing traps at high densities.} Surprisingly, at low densities, the Sokoban walker undergoes self-trapping in which it actively generates its own trap leading to a dynamical localization.  This dynamical localization prevents the walker from percolating to infinity at low densities. We will characterize the trapping crossover through the nonmonotonic dependence of the average trap size on the obstacle density $\rho$, see Fig.~\ref{fig-2d-area}. \bluew{Once again, this behavior is independent of the specific microscopic rules of the Sokoban model and also emerges in its variants featuring different pushing rules (although the value of $\rho _*$ depends on the rule).}

\textit{Survival probability in one dimension:}
In this case, the trapping behavior is determined by the positions of the second obstacles on either side of the walker, which itself is initially at the origin. The walker can push the first obstacle until it is adjacent to the second obstacle. Consequently, it is confined inside the interval $[-L_1, L_2]$ with $L_1 = \left(Y_{O_{2}}^{-} -2\right)$ and $L_2 = \left( Y_{O_{2}}^{+}-2 \right)$. Here, $-Y_{O_{2}}^{-}$ and $Y_{O_{2}}^{+}$ are the second obstacle positions. Given this, the walker will become trapped once it has visited both interval boundaries, $-L_1$ and $L_2$, at least once. Let $Q \left( n|L_1, L_2 \right)$ denote the survival probability that the walker did not interact with \emph{both} of the boundaries until time $n$.

Our strategy is to first construct a renewal framework to calculate $Q \left( n|L_1, L_2 \right)$, and then average it over the probabilities of $L_1$ and $L_2$ to evaluate the unconditional survival probability $S(n) = \langle Q \left( n|L_1, L_2 \right) \rangle _{L_1, L_2} $. In \cite{SinghSM}, we derive that in the diffusive limit, where both $n$ and $L$ are large but their ratio $n/L^2$ remains finite [with $L = (L_1 + L_2)$], the conditional survival probability reads
\begin{equation}
\scalebox{0.9}{$
\begin{split}
Q \left( n|L_1, L_2 \right) \simeq \frac{4}{\pi}~\left[ \sin \left( \frac{\pi  L_1}{2L} \right) + \sin \left( \frac{\pi  L_2}{2L} \right)  \right]~\exp \left(  -\frac{ \pi ^2 n}{8L^2} \right),
\end{split}$} \label{sok-surv-eq-18}
\end{equation}
for $n/L^2 \gg 1$. The probabilities of $L_1$ and $L_2$ are mutually independent and given by $q(L_i) = \rho^2 (L_i+1)\left( 1-\rho \right)^{L_i},\text{ with } i \in \{ 1,2\}$.
Averaging Eq.~\eqref{sok-surv-eq-18}, and applying a saddle-point approximation in the large-$n$ limit, we obtain the unconditional survival probability as
\begin{equation}
\scalebox{0.95}{$
\begin{split}
& S(n) \simeq  ~\mathcal{K}(n) ~\exp \left( -\frac{3 \pi ^{2/3}}{2^{5/3}} \lambda ^{2/3} n^{1/3} \right),~ \text{ as }n \lambda ^2 \gg 1,   \\
& \text{with } \mathcal{K}(n)  \simeq  \frac{2^{\frac{19}{6}} (4-\pi) \rho ^4}{\sqrt{3} \pi ^{\frac{7}{6}} \lambda ^{\frac{5}{3}}}~n^{\frac{7}{6}},~~\lambda = |\ln (1-\rho)|. 
\end{split}$}  \label{sok-surv-eq-35}
\end{equation}
This is the first main result of our Letter. Fig.~\ref{fig-short-n-1d} (left panel) compares this stretched exponential behavior with the numerical simulations for two density values. For both of them, our theoretical expression works well. Comparing Eq.~\eqref{sok-surv-eq-35} with the BVDV formula in Eq.~\eqref{original-eqn}, both survival probabilities, $S(n)$ and $\phi(n)$, are characterized by an exponent proportional to $\lambda ^{2/3} n^{1/3}$. The difference, however, arises in the proportionality constant. It takes the value $ 3 \pi ^{2/3} / 2^{5/3} ~\left( \approx 2.0269 \right) $ for $S(n)$ and $  3 \pi ^{2/3}/2 ~\left( \approx 3.2175 \right) $ for $\phi(n)$. A larger value for $\phi(n)$ means that it decays faster than $S(n)$, since $\phi(n)$ reflects survival against a single obstacle, whereas $S(n)$ accounts for survival against getting caged, a constraint involving multiple obstacles. Another difference is in the prefactor $\mathcal{K}(n)$ in Eq.~\eqref{sok-surv-eq-35}. While it scales as $\sim n^{7/6}$ for $S(n) $, the scaling is $\sim n^{1/2}$ for $\phi(n)$ \citep{Anlauf1984}.

\textit{Moderate-time behavior:} Let us now explore how they compare at small to moderate times. Recall that our long-time $S(n)$ derivation was based on Eq.~\eqref{sok-surv-eq-18} valid for $n/L^2$ large. For small values of $n/L^2$ (with moderate $n$), we, however, obtain a different $Q \left( n|L_1, L_2 \right)$ and averaging it over $L_1$ and $L_2$, we find that the unconditional survival probability behaves as $S(n) \simeq \left( 1- n^2 \rho ^4 / 8 \right)$ for $n \rho ^2 \ll 1$. \\
\indent
Interestingly, this moderate-time behavior can be contrasted with that of $\phi(n)$ in the reactive trapping problem, where the Rosenstock approximation is known to be accurate \cite{Balagurov1974, Rosenstock, Anlauf1984,zumofen1983}. Within this approximation, the survival probability behaves as $\phi(n) \simeq \left( 1 - \sqrt{8 n \rho ^2 / \pi} \right) $ in one dimension. This is qualitatively different from the quadratic decay of $S(n)$ observed here. This highlights how, despite similarities in long-time scaling, the short to moderate time dynamics of reactive and caging-based trapping can be significantly different.

\indent
$N_{\rm P}$-\textit{Sokoban model:} We now examine how the survival probability is affected by variations in the pushing strength of the Sokoban walker. For this, we consider the most general case of $N_{\rm P}$-Sokoban model in which the walker is capable of pushing up to an arbitrary $N_{\rm P}$ number of obstacles. The case $N_{\rm P} = 1$ corresponds to the original Sokoban dynamics as discussed above. For the general case in one dimension, we find that the survival probability is characterized by a large-deviation function \cite{TOUCHETTE20091, LDT-2, LDT-3} in the joint limit $N_{\rm P} \to \infty , n \to \infty$ while keeping the ratio $\omega = n/ \left( N_{\rm P} \right)^3 $ fixed \cite{SinghSM} 
\begin{align}
\lim_{\substack{N_{\rm P} \to \infty , n \to \infty \\    \omega=n/ \left( N_{\rm P} \right)^3  \text{ fixed}         }} \frac{-\ln S \left( n \right)}{N_{\rm P}} = \mathcal{I}\left( \omega \right). \label{gen-LDF-eq-6}
\end{align}
The expression of the rate function reads as
\begin{align}
\mathcal{I}(\omega) = & \frac{\pi ^2 \omega}{32 z(\omega)^2}+2 \left[ z(\omega) \ln z(\omega)-(1+z(\omega))\ln (1+z(\omega)) \right. \nonumber \\
&~~~~~~~~~\left. -z(\omega) \ln(1-\rho)-\ln \rho \right],\label{gen-LDF-eq-9}
\end{align}
where $z(\omega)$ is determined by $\ln \left[ z / (1-z)(1-\rho) \right] = \pi^2 \omega / 32 z^3$.
The idea is to solve this equation numerically and substitute the resulting $z(\omega)$ in Eq.~\eqref{gen-LDF-eq-9} to obtain the rate function $\mathcal{I}(\omega)$. 

The equation above can be analytically solved in different asymptotic limits of $\omega$, yielding the rate function $\mathcal{I}(\omega)$
\begin{align}
\mathcal{I}( \omega) \simeq
\begin{cases}
\frac{\pi^2 \rho ^2}{32(1-\rho)^2}\omega, & ~~~\text{for }  \omega \to 0, \\[4pt]
\frac{3 \pi ^{2/3} \lambda ^{2/3}}{2^{5/3}}~ \omega^{1/3}, &~~~ \text{for }  \omega \to \infty.
\end{cases}
\label{main-result-eq-4}
\end{align} 
Plugging this in Eq.~\eqref{gen-LDF-eq-6}, we find that the survival probability exhibits a crossover from an exponential decay at moderate times to the stretched-exponential decay at large times. \bluew{Interestingly, the stretched-exponential form is independent of $N_{\rm P}$ and is same as Eq.~\eqref{sok-surv-eq-35}. This independence is also shown in the Supplementary Material (SM) using numerical simulations for different values of $N_{\rm P}$ \cite{SMSingh2}. Thus, the long-time survival probability is described by the BVDV formula and is independent of the number of obstacles the walker can push as long as this number is smaller than the system size (implying the limited pushing ability).} This universality stems from the fact the large-time decay is determined by rare events with large voids not shaped by the Sokoban walker.\\

\textit{Survival probability in two dimensions:} Inspired by the matching exponents appearing in $S(n)$ and the BVDV formula in one dimension, we now extend our analysis to the two-dimensional Sokoban model. In this case, analytical treatment, however, proves considerably more challenging and we will resort to numerical simulations.

Rescaling time by the average trapping time $\eta = n  \big/ \langle n_{\rm T} \rangle$, we plot survival probability $S(\eta)$ as a function of $\eta$ in Fig.~\ref{fig-short-n-1d} (right panel). Our key finding is that $S(\eta)$ exhibits a stretched-exponential relaxation at long times, albeit characterized by a different stretch exponent than in one dimension. By fitting the simulation data, we find
\begin{align}
S(\eta) \sim \exp \left(  -f(\rho) \eta ^{1/2} \right),~~\text{for }\eta \gg 1. \label{ahbvci}
\end{align}
This is shown by dashed lines in the figure for two representative density values. 
Based on this fit, we obtain $f(\rho) \sim \langle n_{\rm T} \rangle ^{1/3}$. Interestingly, the stretch exponent $1/2$ in Eq.~\eqref{ahbvci} is exactly same as in the BVDV formula in Eq.~\eqref{original-eqn}. \bluew{Moreover, it remains robust for different pushing dynamics. To examine this, we have studied a variant of the Sokoban model featuring a different pushing mechanism in SM \cite{SMSingh2}. Despite this change in the local microscopic rules, the emergent survival probability still  belongs to the BVDV universality class.}\\

\noindent
\textit{Trapping crossover in two dimensions:} We next analyze the average trap size $\langle A_{\rm T} \rangle$  as a function of $\rho$, which reveals a dynamical crossover governing the trapping behavior. As shown in Fig.~\ref{fig-2d-area}, numerical simulations demonstrate that $\langle A_{\rm T} \rangle$ exhibits a nonmonotonic dependence on $\rho$. Starting from densities close to unity, the mean trap size starts to increase on decreasing the density. This growth continues up to a characteristic density $\rho _* \approx 0.55$, at which point $\langle A_{\rm T} \rangle$  achieves its maximum value. Below this point, $\rho < \rho _*$, we find a turnover behavior in which $\langle A_{\rm T} \rangle$ decreases with decreasing $\rho$. This turnover signals a fundamental shift in the low-density physics, where, despite an increase in the available space for motion, the trap size continues to decrease. We now explain that this shift is related to the emergence of a new trapping mechanism, which is responsible for the loss of percolation.


When $\rho =1$, all sites, except the origin, are occupied by obstacles. The origin, however, contains the walker and it is caged at its initial position. This yields $\langle A_{\rm T} \rangle = 1$. On slightly decreasing $\rho$, a void of vacant sites may emerge and initially surround the walker. Moreover, the walker can enlarge this void by pushing the nearby obstacles. This leads to an increase in the typical trap size $\langle A_{\rm T} \rangle $. As $\rho$ decreases further, the surrounding void grows larger, resulting in a continued increase in $\langle A_{\rm T} \rangle$. This trend, however, continues till the density $\rho _*$. Below this point, $\rho < \rho _*$, a qualitatively different mechanism emerges due to the pushing dynamics. Even when large voids are available for motion, the walker may encounter a rare high-density region containing a manipulable arrangement of obstacles. This region, through successive interactions between walker and obstacles, can be gradually reorganized into a confining structure. This leads to the formation of a localized trap that is not pre-existing but dynamically constructed by the walker itself. As $\rho$ continues to decrease, there are fewer manipulable obstacles giving rise to a smaller trap size. This results in a turnover to decreasing trap size $\langle A_{\rm T} \rangle$, making the overall relation nonmonotonic. This observation, therefore, signals a change in the trapping dynamics from pre-existing traps at high-density to dynamical localization due to the self-trapping mechanism at low-densities. This self-trapping mechanism is also related to the lack of the percolation transition, as eventually the motion of the walker is dynamically localized and it cannot percolate to infinity. Furthermore, we demonstrate in the SM that this turnover and its features remain qualitatively the same even for variants of the Sokoban model with different microscopic pushing rules \cite{SMSingh2}. This further reinforces a universality in the trapping behavior of Sokoban-type walkers.

\blueww{\textit{Other caging studies:} Caging phenomena have been extensively studied across a variety of nonequilibrium settings. In dense active--passive mixtures, an active particle can become temporarily trapped due to the interplay of its self-propulsion and the slow structural relaxation of the surrounding particles \cite{NEB-1,NEB-2,NEB-3}. This results in reversible caging, where the particle is transiently caged for a long time, yet thermal or active fluctuations can eventually enable escape and restore mobility. By contrast, the Sokoban model represents an irreversible caging:  a single active tracer moves through an effectively immobile bath that it irreversibly reshapes. Each displacement reorganizes the surrounding obstacles, progressively shrinking the accessible configuration space until a terminal ``final-cage'' configuration is reached, in which motion is completely arrested within this cage. This irreversible confinement also connects Sokoban dynamics to other nonequilibrium systems that exhibit irreversible kinetic freezing, such as driven and disordered systems approaching absorbing transitions \cite{NEB-4,NEB-5}. However, unlike these systems, where arrest is associated with a sharp absorbing-state transition, the Sokoban model instead displays a smooth crossover between two distinct trapping mechanisms.}

\indent
\textit{Conclusion:} We have investigated the trapping behavior of a Sokoban walker that has a limited ability to modify its environment by pushing a few obstacles. We showed that the survival probability $S(n)$ that the walker has not yet been trapped by time $n$ exhibits a stretched-exponential relaxation in both one and two dimensions, with exponents scaling as $n^{1/3}$ and $n^{1/2}$ respectively. Both of them match with the corresponding exponent in the BVDV formula. However, the accompanying prefactors both within and outside the exponential differ significantly from those in the BVDV framework. Furthermore, the small to moderate time behavior for $S(n)$ is different from that of $\phi(n)$, for which the Rosenstock approximation works well \cite{Rosenstock, Anlauf1984,zumofen1983}. \\
\indent
\blueww{After this, using the average trap size as a proxy, we identified a dynamical trapping crossover that replaces the classical percolation transition in the Sokoban model. The crossover occurs at a density $\rho_* \approx 0.55$ and separates two distinct regimes: a self-trapping regime at low density, where the walker dynamically constructs its own trap, and a pre-existing trapping regime at high density, where confinement arises from the initial arrangement of obstacles.} \\
\indent
\bluew{Notably, both the BVDV behavior and the trapping crossover originate solely from the limited-pushing dynamics, and is independent of the precise microscopic details of the model. This is shown in the SM in both one and two dimensions \cite{SMSingh2}. This robustness makes the phenomenon amenable to experimental verification. In particular, the bristle-robot setup is well-suited for this validation \cite{expt-1}. While this setup, unlike our models, deals with continuous space, caging can still be defined operationally by dividing the arena into discrete grids and inspecting the number of distinct grid cells visited. Saturation of this number signals caging. Another commonly used diagnostic of caging in glassy and biological systems is the time-averaged mean-squared displacement \cite{CT-2007,CT-2020, CT-2022}. A plateau in this quantity signals caging, and the plateau height sets the characteristic cage size. Finally, a convex-hull-based criterion can also be used to detect and size cages \cite{CH-2017}. Convex hull gives a measure of the spatial extent of the particle trajectory  \cite{CH-2009, CH-2010}. A drop in the hull area for the Sokoban walker, say below a certain density, can operationally imply caging and the maximum hull diameter will set the cage size.}

\textit{Acknowledgment:} The support of Israel Science Foundation's grant 1614/21 is gratefully acknowledged.


\bibliography{Bib_new}

\newpage
\begin{widetext}
\onecolumngrid
\setcounter{equation}{0}
\renewcommand{\theequation}{S\arabic{equation}}
\setcounter{figure}{0}
\renewcommand{\thefigure}{S\arabic{figure}}
\section*{SUPPLEMENTARY MATERIAL: Sokoban Random Walk: From Environment Reshaping to Trapping Crossover}
\begin{center}
Prashant Singh, David A. Kessler and Eli Barkai\\
\textit{Department of Physics, Bar-Ilan University, Ramat Gan 52900, Israel}
\end{center}

\section{BVDV trapping universality for the one-dimensional $N_{\rm P}$-Sokoban model}
\indent
In the letter, we discussed the $N_{\rm P}$-Sokoban model, which generalizes the original Sokoban model by allowing the random walker to push up to an arbitrary $N_{\rm P}$ number of obstacles in one dimension. We found that for large $n$, the survival probability $S(n)$ has the stretched exponential relaxation
\begin{align}
S(n) \sim \exp \left( -\frac{3 \pi ^{2/3}}{2^{5/3}} \lambda ^{2/3} n^{1/3} \right),~ \text{ as }n \gg N_{\rm P}^3,
\label{SM-eq-1}
\end{align}
indicating the BVDV trapping universality. In Fig.~\ref{fig-SM-1d}, we have compared this relaxation with the numerical simulations for three pushing strengths, namely $N_{\rm P}=2,5,8$, with the obstacle density fixed to $\rho = 0.1$. In all cases, the simulation data shows excellent agreement with Eq.~\eqref{SM-eq-1} at sufficiently large $n$. Therefore, the survival probability $S(n)$ exhibits the BVDV universality regardless of the microscopic pushing rule of the Sokoban walker.
\begin{figure}[b]
    \centering
    \includegraphics[scale=0.3]{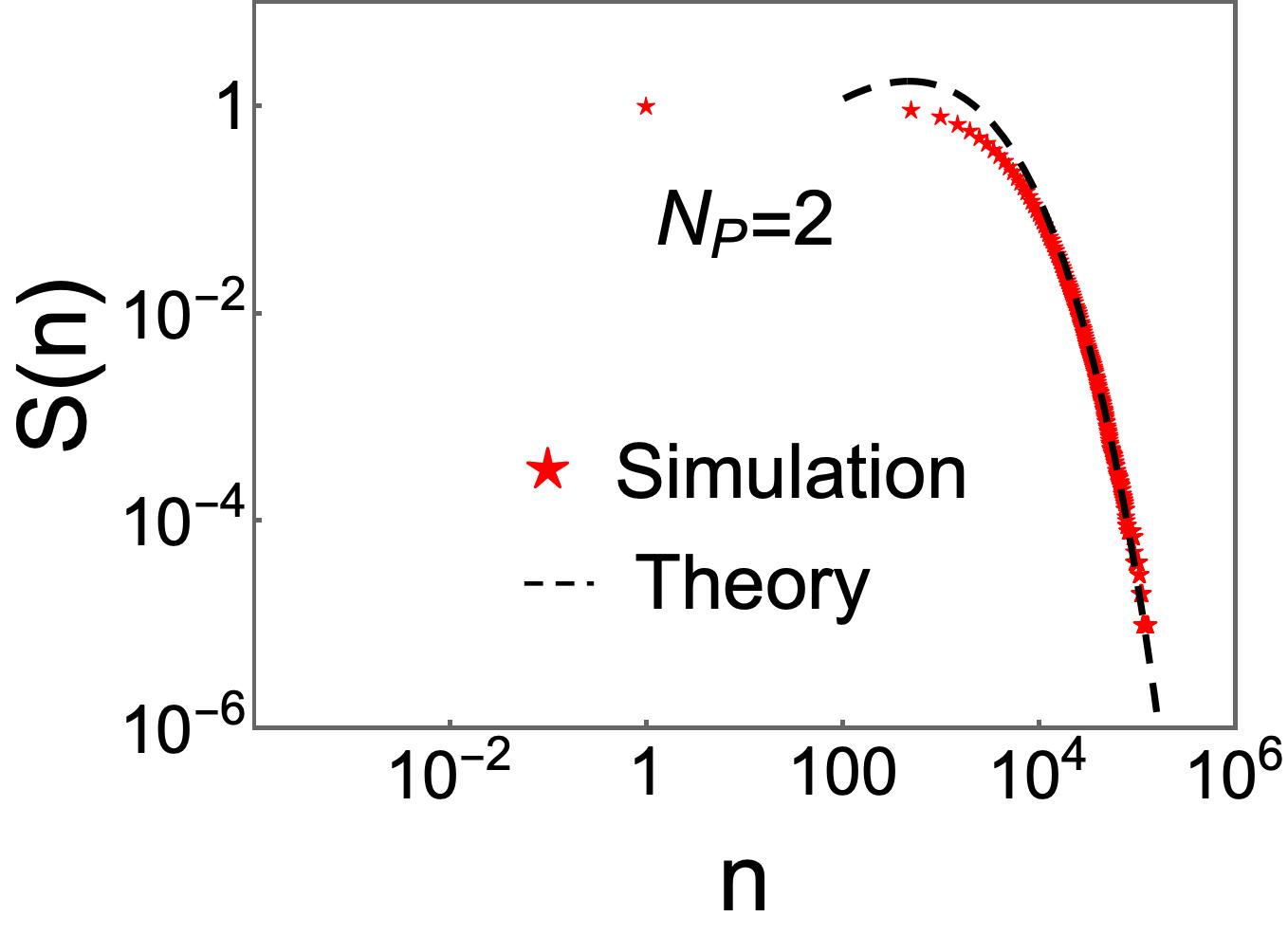}
     \includegraphics[scale=0.3]{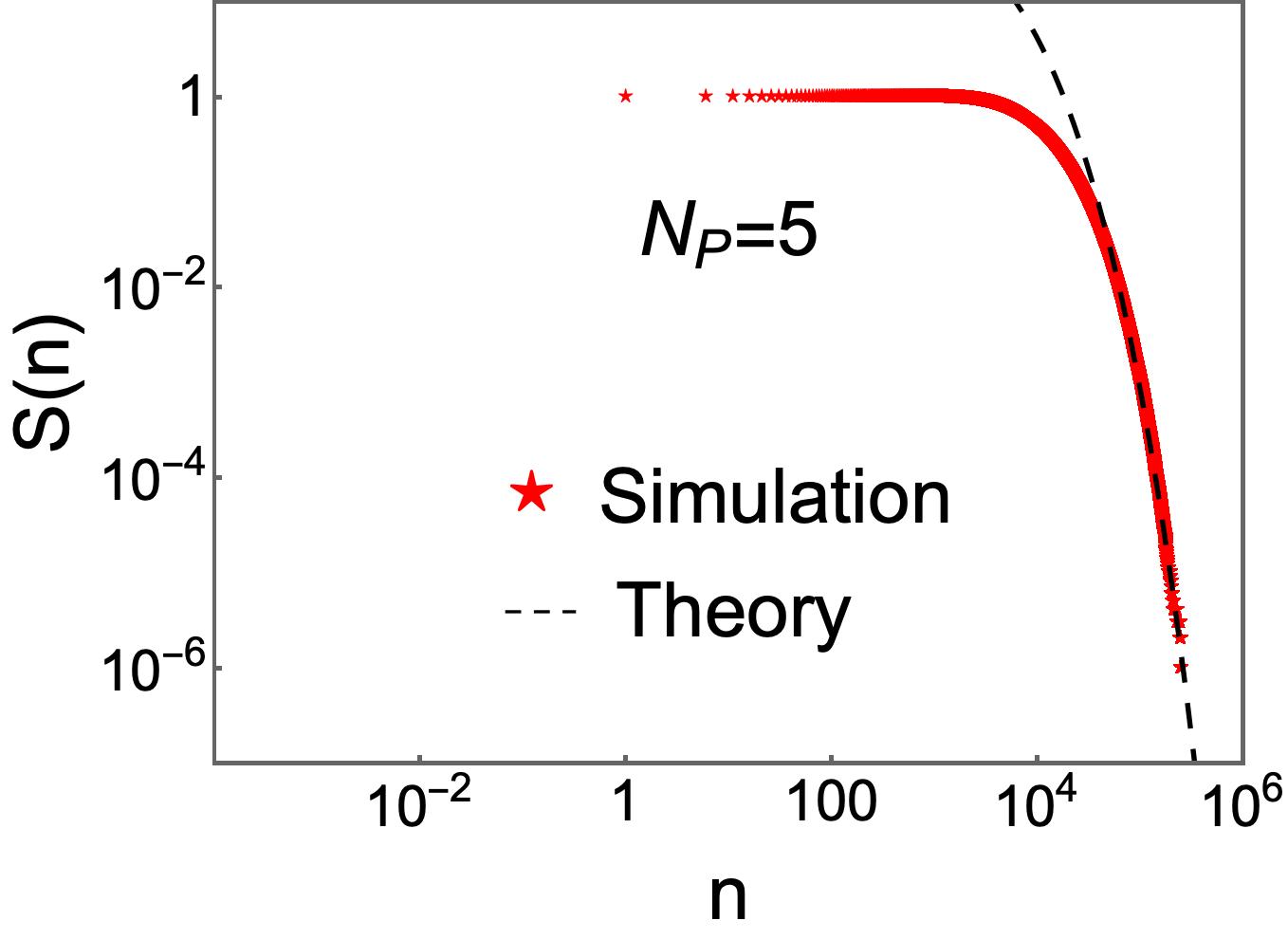}
      \includegraphics[scale=0.3]{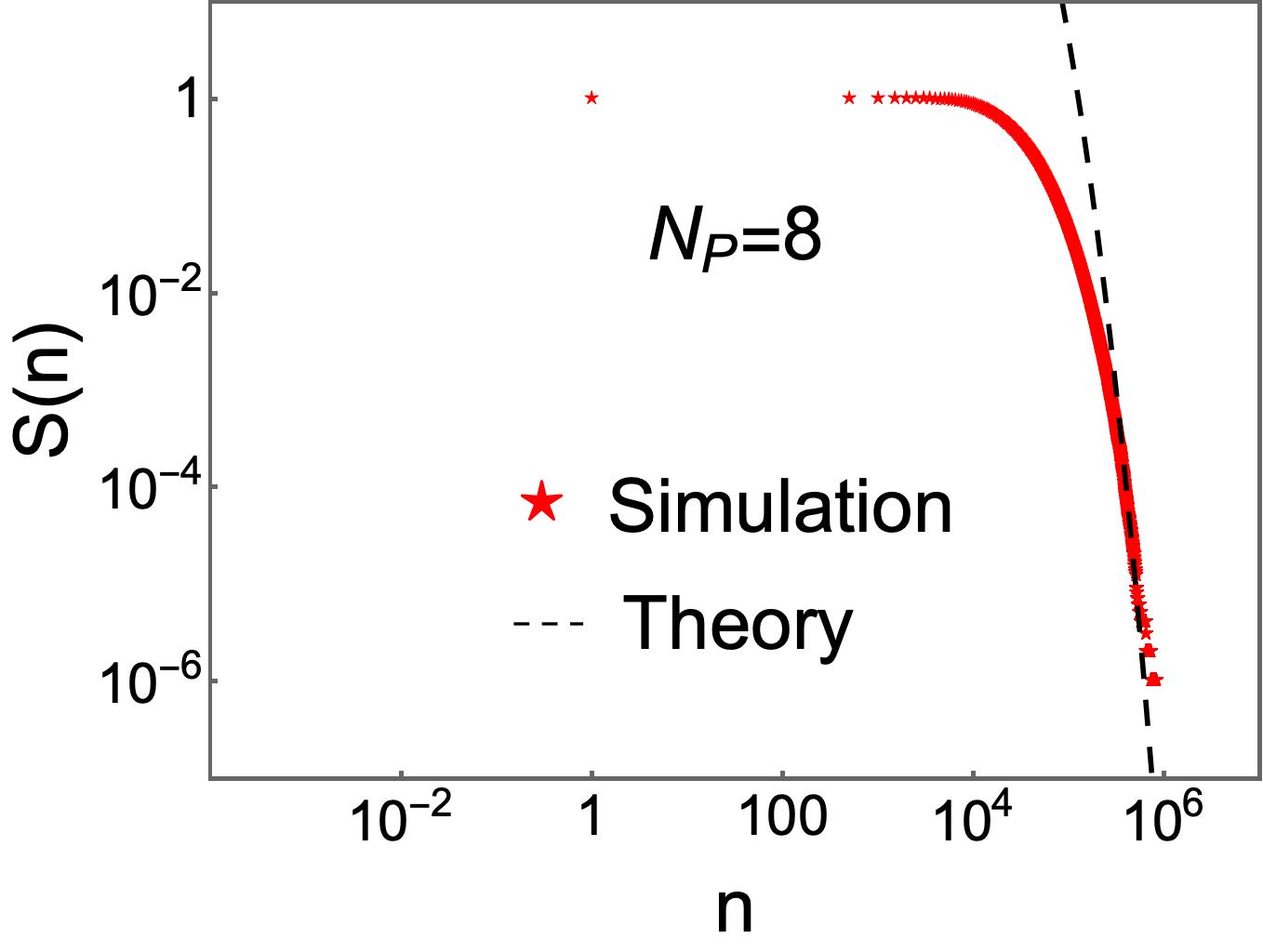}
    \caption{Survival probability $S(n)$ for the one-dimensional $N_{\rm P}$-Sokoban model for three different values of $N_{\rm P}$. The obstacle density is fixed to $\rho = 0.1$ for all three cases. The red symbols denote the simulation data, while the dashed line represents the stretched-exponential form in Eq.~\eqref{SM-eq-1} associated with the BVDV universality.}
    \label{fig-SM-1d}
\end{figure}

\begin{figure}[t]
    \centering
    \includegraphics[scale=0.3]{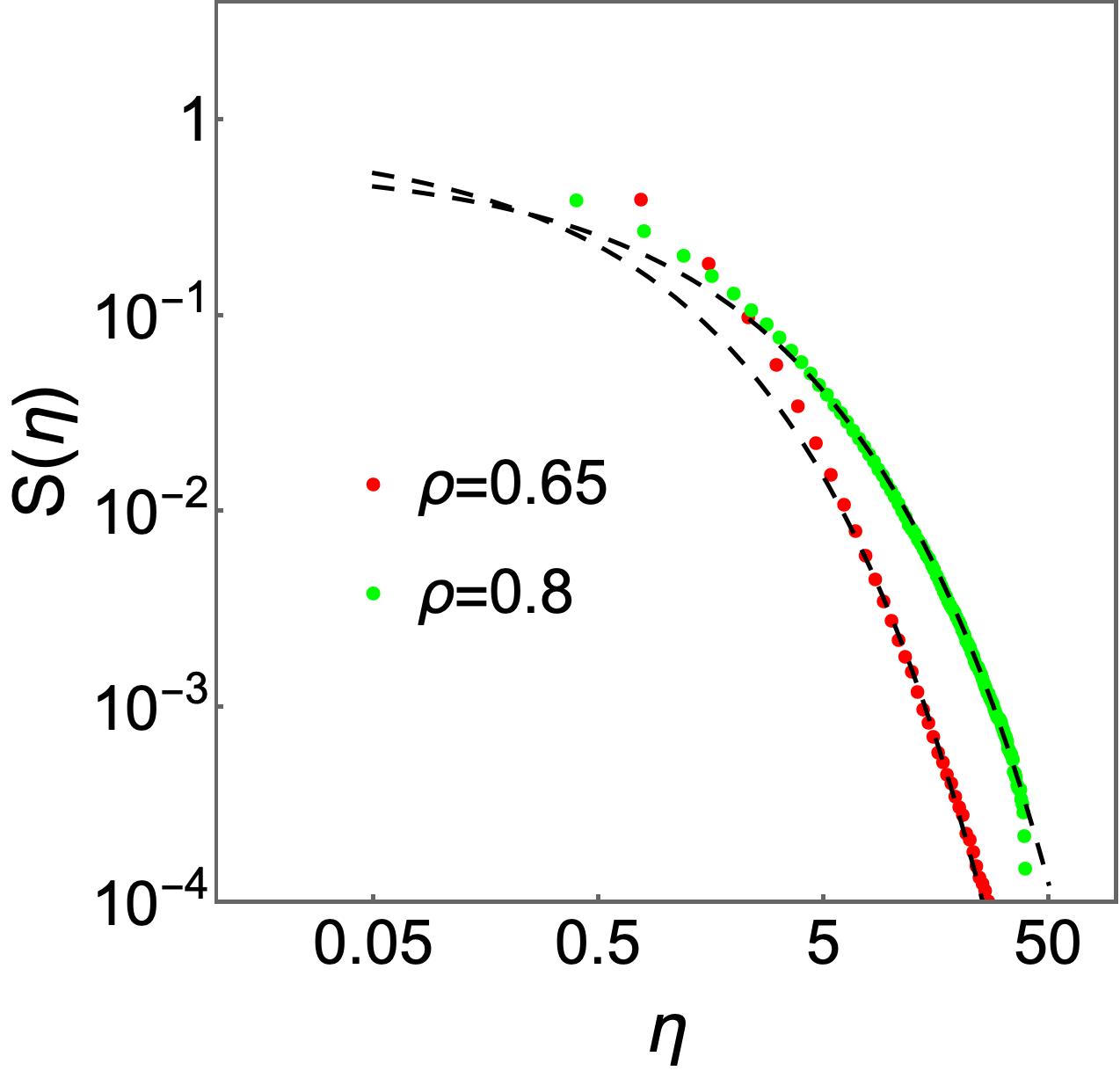} \hspace{0.3 cm}
    \includegraphics[scale=0.285]{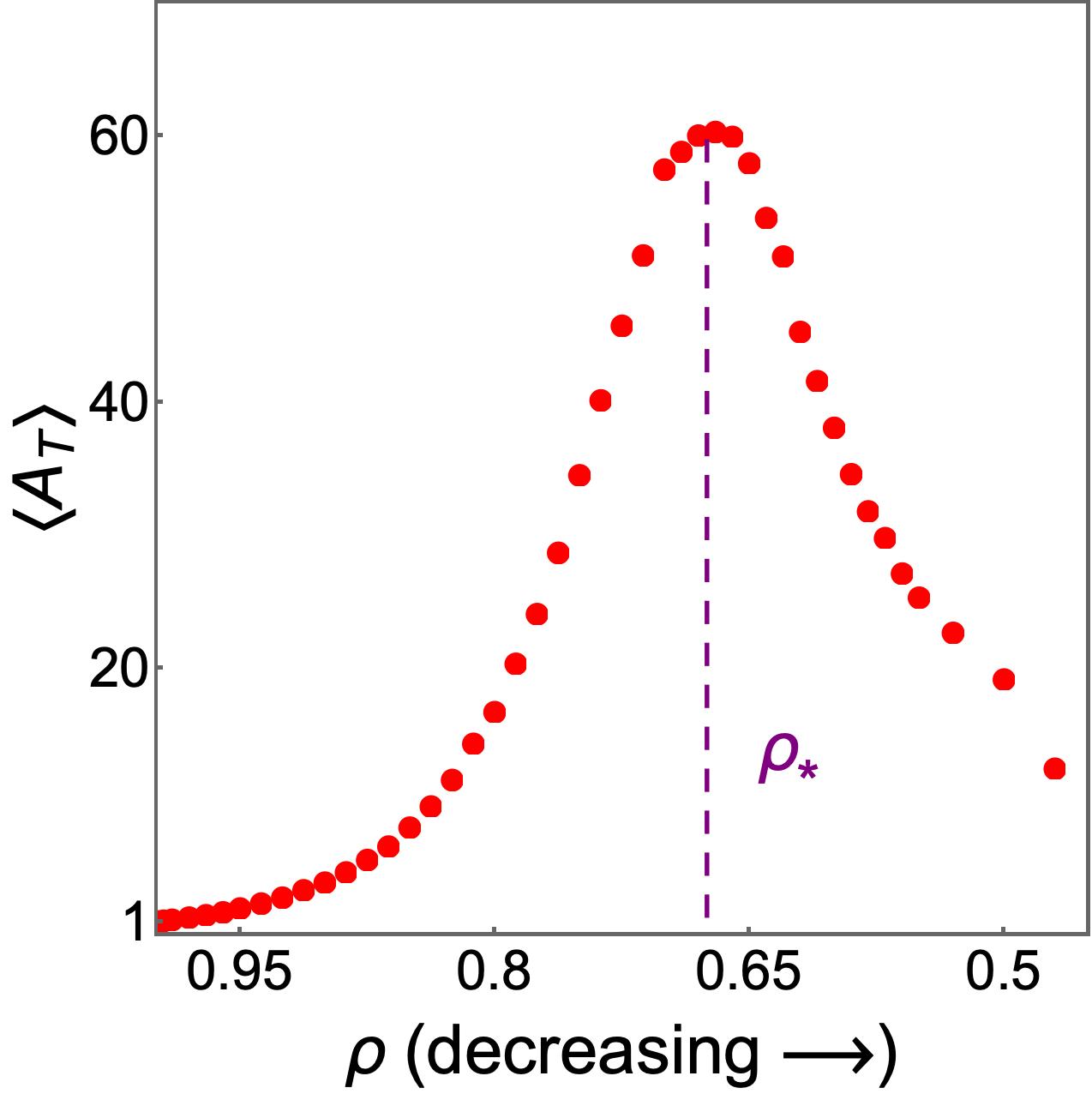}
    \caption{\textit{Left Panel:} Survival probability $S(\eta)$ vs $\eta = n / \langle n_{\rm T} \rangle$ for the two-dimensional generalized Sokoban model for two different values of density. The colored symbols represent the simulation data, while the dashed lines are fit to the data corresponding to BVDV form in Eq.~\eqref{SM-eq-2}. \textit{Right Panel:} Dependence of the mean trap size $\langle A_{\rm T} \rangle$ on obstacle density~$\rho$. The curve is nonmonotonic, with $\langle A_{\rm T} \rangle$ attaining its maximum value of approximately $60$ at $\rho_* \approx 0.675$. This peak reflects the \blueww{crossover} between pre-existing trapping at high densities and self-trapping at lower densities, similar to the behavior observed in the original Sokoban model.}
    \label{fig-SM-2d}
\end{figure}
\section{BVDV trapping universality and trapping crossover for the generalized Sokoban model in two dimensions}
In order to verify the universality of the trapping phenomenon, we examine the generalized version of the original Sokoban model. In this generalized model, an obstacle can be pushed to any of its three neighbouring sites (except the site containing the walker) with equal probability $1/3$, provided that this target site is vacant. This is different than in the original Sokoban model, where the obstacle can be pushed only in the direction of the walker's attempted motion. Here, we will show that changing the microscopic rules does not qualitatively change the main results of the letter, namely the BVDV decay for the survival probability and the trapping crossover.

In the left panel of Fig.~\ref{fig-SM-2d}, we perform extensive numerical simulations to obtain the survival probability $S(\eta)$ as a function of $\eta = n / \langle n_{\rm T} \rangle$, where $\langle n_{\rm T} \rangle$ is the average trapping time. \blueww{We show the simulation result for two obstacle densities, $\rho=0.65$ (red) and $\rho = 0.8$ (green).} For both cases, we find that the long-time decay is captured by
\begin{align}
S(\eta) \sim \exp \left(  -f(\rho) \eta ^{1/2} \right),~~\text{for }\eta \gg 1. \label{SM-eq-2}
\end{align}
with $f(\rho)\sim \langle n_{\rm T} \rangle^{1/10}$. The dashed lines represent the corresponding stretched-exponential fits. These results demonstrate that even in the generalized Sokoban model, where the microscopic pushing rules are modified, the survival probability still belongs to the BVDV trapping universality class.

Furthermore, the right panel of Fig.~\ref{fig-SM-2d} illustrates the dependence of average trap size $\langle A_{\rm T} \rangle$ on $\rho$ for the generalized-Sokoban model. Similar to the original Sokoban dynamics, we see that this dependence of $\langle A_{\rm T} \rangle$ is nonmonotonic and the mean achieves its maximum value of $\langle A_{\rm T} \rangle \approx 60$ at $\rho _* \approx 0.675$. As explained in the letter, this turnover arises due to two different mechanisms governing the trapping dynamics of the walker: a self-trapping mechanism at low density, where the walker becomes dynamically localized within a self-formed trap, and a pre-existing trapping regime at high density, where confinement arises from the initial arrangement of obstacles. To summarize, the results in the right panel of Fig.~\ref{fig-SM-2d} demonstrate that the trapping crossover behavior is robust against changes in the microscopic pushing rules of the Sokoban model. While the precise value of the crossover density $\rho_*$ changes, the qualitative shape remains unchanged.
\end{widetext}
\end{document}